# USING SMARTPHONES AND TABLETS AS EXPERIMENTAL TOOLS IN THE PHYSICS CLASSROOM: EFFECTS ON LEARNING AND MOTIVATION

Authors: A. Gasparini[a)], F. Stern[a)], M. Delaval[b)], A. Müller[a),c)]
Affiliations:
a)  Institute of Teacher Education, University of Geneva, Switzerland
b)  Univ. Lille, ULR 4072 – PSITEC – Psychologie: Interactions, Temps, Emotions, Cognition, Lille, France
c)  Department of Physics, University of Geneva, Switzerland

**Abstract**

Tablets and smartphones provide compact and relatively inexpensive tools for measuring many physical quantities, and numerous applications of these "mobile devices as experimental tools" (MDETs) have been published over the last decade, covering a broad range of topics. According to the literature, MDETs can offer educational benefits by creating authentic, real-life contexts for physics learning, enhancing student motivation through the use of familiar technology, and supporting cognitive processes by providing multiple representations (videos, tables, graphs) of physical phenomena. On the other hand, serious concerns have been raised about potential distractions and cognitive overload. Regarding these conflicting perspectives, few empirical studies on the impact of MDETs in real classroom settings of regular, full-length physics courses are available, and – to the best of our knowledge – none focus on a non-specialized high-school target group.

The purpose of this contribution is to present a study of a mechanics course in such a setting, addressing the tight curricular, material, and practical constraints inherent to it. A quasi-experimental pre/post design comparing a treatment group using MDETs and a control group without (same content, lesson plan, and teacher) was used. The 19-week teaching sequence focused on conceptual learning and motivational outcomes (e.g., interest, self-concept, and curiosity), controlled by several predictor variables.

Findings reveal substantial pre/post learning gains for both groups (Cohen's $d = 0.9$) and small gains for perceived relation to reality ($d = 0.29$). However, no significant differences between treatments were found, indicating that the use of MDETs does not outperform conventional approaches under the given constraints. On the other hand, no evidence of negative effects such as distraction or cognitive overload was observed. Moreover, little to no interactions with predictors such as gender or prior knowledge were found.

In conclusion, MDETs show considerable potential as a viable and effective option for integrating technology into physics teaching, offering learning outcomes comparable to those of successful conventional teaching, but not better. Furthermore, no evidence of negative effects from distraction or cognitive overload was found, nor of strong interactions with gender and other predictors. The MDETs approach thus appears suitable for both sexes and different kinds of learners. This leads to suggestions for future utilization of MDETs as a component of physics education, which are discussed as potential perspectives.

# 1 Introduction

Tablets and smartphones are easy to handle and widespread among people of all generations; their use as learning tools in science classes is nowadays a reality [1] [2]. The built-in sensors provide a new kind of compact and relatively inexpensive instruments for many physical quantities. In physics education, a broad range of applications of these mobile devices as experimental tools (MDETs[1]) have been published since about a decade, covering mechanics [3] [4], acoustics and waves [5] [6] [7] [8] [9] [10], optics [11] [12], magnetism [13] [14], radioactivity [15] [16] and other fields. For reviews see [1] and [17]; for a recent broad collection see [2]. Phyphox (Physical Phone Experiments) is a highly versatile, user-friendly open-source software for sensor control, readout and data export for use with MDETs [18]. A first advantage of using MDETs in physics education is the possibility to perform relatively (with respect to traditional classroom instruments) effortless and at the same time precise measurements, supporting a real-time interpretation of the data, both in the classroom or in the everyday-life situations. This makes it possible for learners to learn science in a "topical" context (a situation) that is perceived as more authentic than traditional sciences courses, e.g. a vertical jump as an application of Newtonian mechanics and energy conservation [4]; see also the numerous examples cited above. Moreover, as these devices are part of everyday life of learners, their utilization during physics lessons may provide an authentic "material" context (an instrument) for them [19]. This authenticity matches the framework of Context-Based Science Education [20] [21] based on theoretical and empirical arguments that providing meaningful, real-life contexts can have a positive impact on motivation and learning (see 2.1). However, a positive effect on motivation implies only an indirect effect for learning, and other existing research implies conflicting arguments for direct effects of MDETs regarding physics learning (see sect. 2.2 and 2.3). Moreover, the empirical evidence of the direct effects of smartphone use on physics learning seems to be inconclusive at present [22]. On one hand, they represent powerful instruments whose apps can easily and straightforwardly produce tables, diagrams, and many other functions that – in a traditional course – have to be produced by the learners themselves. This can lead to improved learning, as already shown for other ICT (Information and Communications Technology) applications to physics education [23]. Yet other studies also indicated negative effects of ICT on learning, in particular distraction and cognitive overload, impeding learning [24] [25] [26].

Specifically regarding MDETs, many experiments have been proposed in the last decade (see references above), but only few empirical studies about their educational impact exist, which were either carried out in single lab sessions (about three hours; [19]), among advanced learners ([27] - [30]), or both [31]. The goal of our investigation is thus to examine the effects of the use of MDETs in physics education, taking into account and completing the previous results. The main differences with respect to previous studies are the systematic integration into a whole learning sequence in regular physics classroom, at high school level, and the duration of the intervention (36 lessons of 45').

# 2 Research background

Prior studies have established a well-developed research background for the use of MDETs in physics education, integrating elements of context-based science education, cognitive-affective theory of multimedia learning, and other elements from research in science education and educational science. Two further well-established areas of physics education research on which we draw are conceptual learning, and the connection between mathematics and physics learning. We review each of these strands of research with a focus on elements relevant for the present study.

---

[1] To the best of our knowledge, the acronym M(D)ET was coined by Kuhn & Vogt (2015), together with similar ones (e.g. NET = New Experimental Tool).

## 2.1 Context-based science learning

As stated above, a remarkable advantage of using MDETs in science classes is the possibility to perform new types of activities, creating new situations that would not be possible using a traditional lab setup, whether inside or outside the classroom. Thus, new contexts for science learning can be conceived, closer to the everyday life of the students and to their real interest. This matches the theoretical framework of Context Based Science Education (CBSE; [20] [21]), positing that science learning related to actual, real(istic), genuine contexts and experiences learners are supposed to encounter can foster their motivational and cognitive activation, and thus lead to improved attitudes to and understanding of science. Indeed such positive effects on attitudes/motivation (with large effect sizes) and learning (with medium effect sizes) have been revealed in some empirical studies ([20] [32] [33]). The approach of CBSE is also strongly advocated by PISA (Program for International Student Assessment), see e.g. the statement of the importance of learning activities "that could be part of the actual experience or practice of the participant in some real-world setting [and] places most value on tasks that could be encountered in a variety of real-world situations" [34], or by the in-depth analysis of Fensham [21]: "[R]eal world contexts have [...] been a central feature of the OECD's PISA project for the assessment of scientific literacy among young people". This focus aligns well with the importance of real-life phenomena and real-life problems within the "Next Generation Science Standards" in the United States [35].

Despite encouraging empirical evidence and the recommendations in favor of context-based science learning, a main obstacle to its widespread integration into the practice of science teaching comes from the difficulties encountered to prepare such courses in the long term, requiring a substantial adaptation of the existing pedagogical means in relation to the actual curricula [36]. In other words, it is often difficult for teachers to reconcile the educational potential of CBSE with practical and curricular requirements, particularly over the long term.

MDETs constitute a new educational tool allowing to reconcile the practice of context-based activities with existing educational constraints and settings. As argued by Hochberg *et al.* [19], they can provide a learning context in two different and related ways: First, as mentioned above, the flexibility and the "hands-on" format of the MDETs activities opens up the possibility of new contexts for science learning, perceived as relevant to the everyday life of learners, whether inside or outside the school. For example, activities whose context is sport or a hobby practiced by the student: this is called a *topical context*. Topical contexts are known to be able to foster interest and learning in science [20]. Second, smartphones and tablets are widespread and often have an important emotional value to young people: therefore, the use of these "hands-on" and attractive devices as measuring instruments during the physics lessons is an element of authenticity in itself, constituting what is called a *material context* (the device) [19]. Moreover, this exploits the attractiveness of new technologies to strengthen the interest of students in the studied topics (as shown by Swarat *et al.*, [37]. However, an interest based only on the material context linked to mobile devices needs to be maintained in order to be transferred to the subject matter on a long-term basis, which is unlikely in the context of one-off activities, but rather requires a prolonged use of MDETs.

Among the two types of context (topical, material) possible with MDETs, the topical context is likely to promote a positive attitude towards sciences in a more direct way, because it is based on a more explicit link between the studied discipline and the center of interest of learners. Nevertheless, the effect of the topical context is reinforced if the material context is also present. Carrying out activities that combine both contexts, material and topical, is a promising approach to make the most of the potential of MDETs in sciences lessons [19]. A main purpose of the present work is to implement and study MDETs in the real classroom setting of a regular high school mechanics course.

Focusing on motivational outcomes and following the above reasoning, two important sub-dimensions of motivation might directly be impacted by the use of MDETs in the physics lessons and will be object of our investigation: the learners' *perception of the relation to reality* as well as the *interest in relation to the learning subject*. *Perceived relation to reality* reflects the extent to which the subjects treated in the course appear relevant and linked to the daily life of the learners and therefore it corresponds also to a measure of the perceived authenticity. *Interest* is defined as "a psychological state of engaging [...] that occurs and develops during interactions between persons and their environment as objects, topics, contents or ideas; interest is characterized by focused attention and a motivational reaction, such as enjoyment" [38] [39] [40]. Another key internal state of motivation in our investigation is the construct of *curiosity as a state*, that is the curiosity toward the subject of learning, defined as curiosity as "a desire for new information or experience that includes a trigger, a reaction to that trigger, and a resolution, which can be satisfactory or unsatisfactory". Indeed, if the curiosity of a learner is triggered for example by a stimulating context provided by an activity in the physics course, this can be the source of her or his wish to seek new information and to engage in a further learning, renewing in turn the initial state of curiosity [41].

In addition, the relative ease of access to the data and their representations makes it possible to minimize the effort of the learner, provided that she or he masters the functionalities of the apps used during the activities. Thus, the perceived competence of a learner to quickly, easily and successfully conduct MDETs activities can be increased and this, in turn, can be beneficial both for the motivation and learning. This aspect is commonly understood in terms of *(situational) self-concept*, i.e. "a mental representation of oneself, including a collection of cognitive beliefs about oneself formed through experience and feedback from the environment" [42] [43]. We thus include *situational self-concept* experienced during a learning sequence as a dependent variable in our study.

## 2.2 Cognitive (-affective) theory of multimedia learning

In section 2.1, we explained how the creation of a meaningful context can foster motivation in science learning. Despite the widespread belief that a positive effect on motivation can lead to a positive effect on learning, meta-analysis have only shown a moderate correlation (0.3-0.4) between motivation and learning [44] [45]. As a consequence, based solely on the effects that MDETs can have on motivation, one can only expect moderate *indirect* effects on learning. However, further arguments and empirical observations exist, regarding possible *direct* effects of MDETs on learning. These are framed in the *cognitive-affective[2] theory of learning with media* (CATML; [46] [47]). This theory expands on the *cognitive theory of multimedia learning* (CTML; [48] [49] [50]) by integrating affective factors in the learning process, especially regarding attention allocation and cognitive engagement. Common to both theories is a strong foundation in cognitive load theory ([51] and theories of multiple representations (dual coding theory [52] and extensions to further modalities [46]).

C(A)TML is based on several assumptions derived from prior research [47] [53] [54] : (i) humans process different information modalities through different channels; (ii) processing of current information occurs in short term (or working) memory with limited capacity for each channel; (iii) storage takes place in long-term memory, an active, evolving system which holds and reproduces information on the basis of various mental processes (structuring, inference mechanisms, emotional tags, etc.); short and long term-memory interact in both directions; (iv) meaningful learning takes place when learners are actively engaged in cognitive processes such as selecting, organizing (both within and between channels), and integrating new and existing information (retrieved from long-term memory); (v) affective factors can mediate aspects such as attention allocation in the short term memory or active engagement in cognitive processes (related to assumptions (ii) and (iv), respectively)[3]. Assumptions (i) through (iv) are common to both CTML and CATML, while assumption (v) is specific to the latter.

---

[2] `Affective´ is sometimes used as an umbrella term for emotional and motivational dimensions including interest and self-believes [53].

One can see how multiple representations are related to assumptions (i), (ii) and (iv); physics learning, in particular, systematically requires simultaneous understanding e.g. of a real experimental setting, a schematic representation, diagrams, tables, and mathematical expressions ([55] [56] [57]). Cognitive load theory, in turn, comes into play in assumptions (ii) and (iv). Specifically, two aspects are of particular interest here: First, the concept of "element interactivity" as key factor for cognitive load (CL), where an "element" is a concept or a procedure to be learned and "interactivity" represents the degree of linkage between the elements necessary for a given learning task. Learning distinct elements with only a few connections (as simply memorizing the three Newton's laws) is considered a "low element interactivity" task. In contrast, a task has a "strong element interactivity" when there are many connections between the different elements, for example solving complex problems or carrying out physics experiments, where the links between the observations and the theory have to be understood and investigated. With the many multiple representations occurring naturally when learning and practicing physics (see above), element interactivity and thus CL are high. Second, the differentiation between "intrinsic" and "extraneous" cognitive load: *intrinsic cognitive load* is generated by the nature of the learning content and is inherent to the complexity of the learning subject itself (in particular due to high element interactivity, [58]), while *extraneous cognitive load* arises from the presentation of the content to be taught, in particular surface features, and does not contribute to learning. Henceforth, extraneous load should be reduced as much as possible to keep the limited short-term memory capacity available to deal with the intrinsic cognitive load. CATML emphasizes that design and presentation features of media that can lower cognitive load, thus helping to enhance learning outcomes. We will now apply this framework to MDETs as educational approach in the physics classroom.

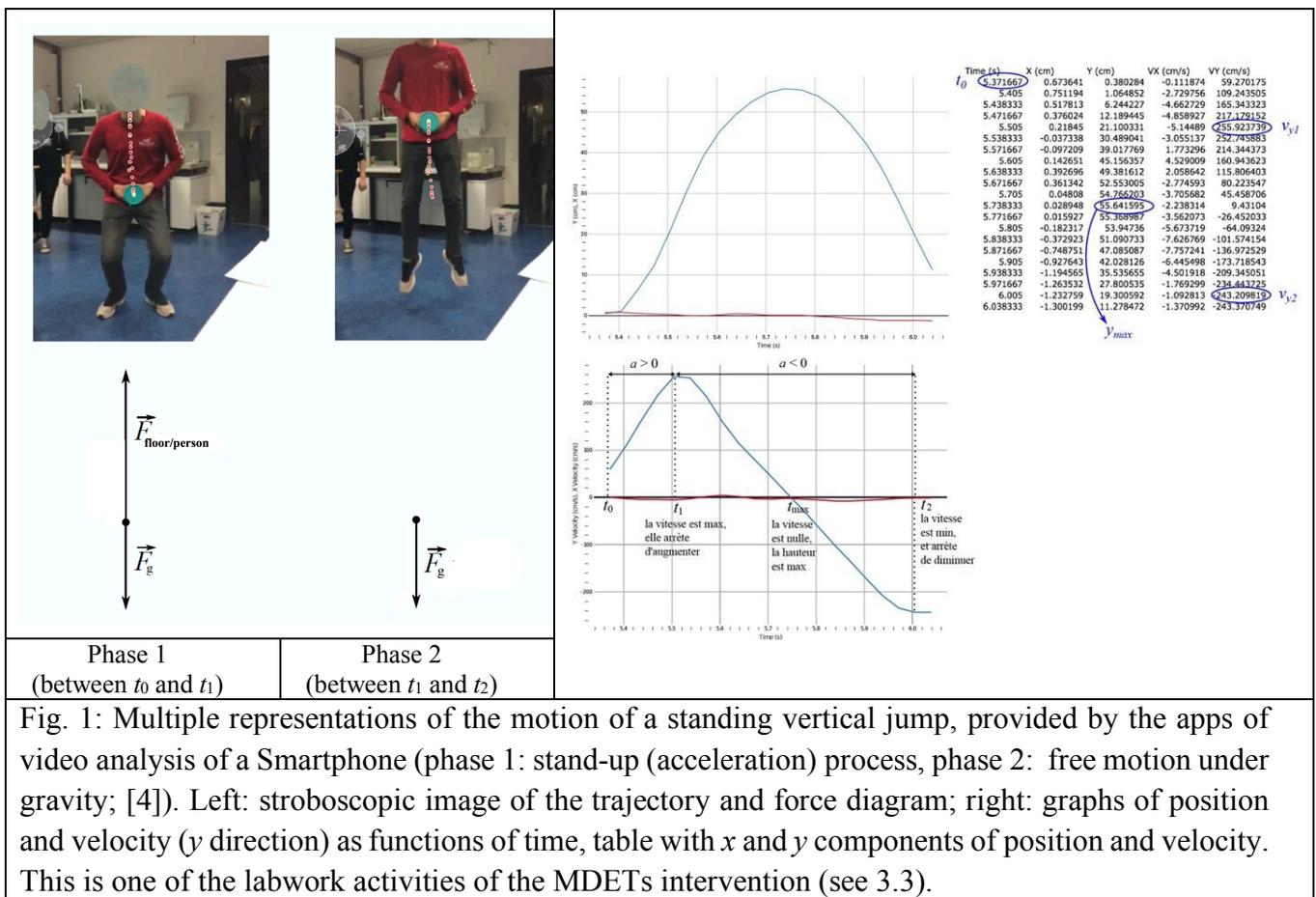

Fig. 1: Multiple representations of the motion of a standing vertical jump, provided by the apps of video analysis of a Smartphone (phase 1: stand-up (acceleration) process, phase 2: free motion under gravity; [4]). Left: stroboscopic image of the trajectory and force diagram; right: graphs of position and velocity (*y* direction) as functions of time, table with *x* and *y* components of position and velocity. This is one of the labwork activities of the MDETs intervention (see 3.3).

---

[3] These mediation factors are probably related to the indirect effects of motivation on learning discussed above. However (to the best of our knowledge), available research has not analyzed this question in detail.

## 2.3 MDETs as educational approach in the physics classroom

MDETs represent powerful instruments with the potential to reduce the extraneous CL during experiments, as the apps can easily and straightforwardly provide multiple representations and many other functions that – in a traditional course – have to be produced by the learners themselves. An example of multiple representations of the motion of the standing vertical jump ("squat jump"), [4]) provided by the apps of video analysis is shown in Fig. 1. These representations can be used for example to calculate the maximum jump height using various foundational relations of Newtonian mechanics. Other examples of similar kind can be found in collections [1] and [2]. These examples demonstrate how a considerable amount and diversity of data on motion experiments can be easily stored and analyzed within the mobile device, depending on learning objectives and level of learner. These possibilities leave more time and attention to interpret, classify and organize relevant information and work out the element interactivity.

Based on its theoretical background and empirical evidence supporting it, C(A)TML has developed specific guidelines and design principles for instructional media [53] [54]. The following principles are of special importance with respect to MDETs [31]:

The principle of contiguity (spatial and temporal) for multiple representations states that (i) spatial proximity helps minimizing visual search processes, which can lead to a reduction in extraneous CL; (ii) temporal proximity helps avoiding additional CL that arises when mental representations need to be held active for longer durations. The above example (Fig. 1) show how the simultaneous display of corresponding and complementary representations possible with MDETs fulfills the requirement of spatial and (by "swiping" rapidly between apps) also of temporal continuity.

The principle of interactivity (or segmenting principle) states that better learning occurs when learners can explore dynamic visualizations with sequence and pace determined by themselves. With MDETS, learners have full control over the different representations (real motion, stroboscopic image, tables, diagrams) and can for instance switch forth and back between a stroboscopic image and the velocity diagram in order to understand their connection (an essential example of element interactivity for the learning of mechanics).

Finally, the positively moderating effect of motivation on attention allocation and active engagement in the learning process with MDETs was already discussed above.

Yet there are also plausible arguments that MDETs might actually not be conducive to learning. First, one might say that the offload of cognitive load in producing multiple representations (e.g. motion diagrams) is in fact counterproductive, since it is essential for learners to engage in this process themselves. Indeed, learning with multiple representations is known to be critical issue in science learning, often creating considerable difficulties [59] [60] [61]. A solid understanding lies above all in the ability to make connections between the different representations and to create coherent connections within each representational format, that is to say to create intra- and inter-representational connections [50] [62] [63]. According to Mayer ([50] [63] [64]), these connections and the consistency between them are essential for effective learning and the positive impact of actively processing different representations and their connections has be shown empirically [57]. This is the reason why caution is necessary when introducing these multiple representations into science learning sequences: learners must have the necessary prerequisites for handling all the representations provided (e.g. in mechanics diagrams for $x(t)$, $v(t)$). Hochberg et al [31] rightly state, that a "blanket statement on how digital media can be meaningfully used in the classroom is not possible", and whether the facilitated use of multiple representations via MDETs is actually such a meaningful use is an open question.

Second, it is important to note that the above benefits are only possible for learners who, beyond sufficient understanding or relevant representations (as just stated) additionally are rapidly able to handle the necessary functionalities on the device. Otherwise their use may turn out to be counterproductive and overload working memory, thus preventing learners from focusing on the content to be learned. Moreover, distracting effects of ICT in general can be observed, if its use in the classroom is not sufficiently structured (see e.g. [65] for a study regarding laptops). Learners can make an inappropriate use of the devices, for instance by processing information which is not correctly connected to the learning purpose, or by just "playing around", which is likely to lead to an increased extraneous CL, and also less attention and less cognitive engagement. On the other hand, several more recent studies did not find such distracting effects of MDETs when used for physics learning, and they rather reported positive effects [19] [27] [31]. In fact, gaining empirical evidence on this interplay of positive and negative effects is a main research objective of this research.

## 2.4 Purpose and research questions

Even though many experiments using MDETs have been proposed in the last decade (see sect. 1), only few empirical studies about their educational impact exist (see Tab. 1). Hochberg et al. ([19] [31]) used MDETs with upper secondary students in a single a learning unit on mechanics (pendulum). The results of their first study have shown higher interest levels in learners of the treatment group (with a small to medium effect size), as well as a higher curiosity related to the studied subjects (small effect size; no differences in learning were observed [19]. In contrast, the second study among learners of a specialized physics course reports significantly higher levels of learning in the treatment group (with medium effect size), while no differences have been observed for motivational outcomes and cognitive load [31]). None of these two studies used a topical context, which leaves unanswered questions regarding the effects of using MDETs when exploiting its full potential, i.e. combining the topical and the material context. In another investigation in an undergraduate mechanics course, lasting a full semester, and with both material and the topical contextualization taken into account, Klein et al. ([27] [28]) obtained medium to large positive effects for conceptual understanding, relation to reality, curiosity, disciplinary authenticity, self-concept, interest, and autonomy, consistent with other work in the same target group [66] [67]. Note that the studies having found positive effects on learning were among learner selections with higher levels of motivation and understanding (specialized high school [31] or undergraduate students [27]).

Today, to the best of our knowledge, there is no study on the use of MDETs in the real classroom setting of a regular mechanics course for upper secondary level students not specializing in physics and covering a whole teaching sequence matched to the official study plan. Note that such a course faces tight constraints, limiting feasible experiments and topical contexts. A main purpose of the present work is to present such an implementation of MDETs, combining material and topical context as far as possible within the given limitations.

The choice of learners not specializing in physics implies a focus on conceptual learning as educational objective, addressing learning difficulties and misconceptions well-known from prior research in mechanics learning ([68]-[71], ch. 14.1.2), including established assessment instruments ([72], sect. III).

Tab. 1: Summary of the main features of recent studies on the effects of MDETs in mechanics courses.
*within the constraints of regular high school mechanics course matched to the official study plan.

| Study | Educational level | Specialized target group | Duration of the intervention | Topical context | Material context |
|---|---|---|---|---|---|
| [19] | Secondary II | No | Single lab session | No | Yes |
| [31] | Secondary II | Yes | Single lab session | No | Yes |
| [27], [66], [67] | Undergraduate | Yes | One semester | Yes | Yes |
| This study | Secondary II | No | One semester | Yes* | Yes |

In this perspective, we produced a series of MDETs activities, which will suitably replace the traditional laboratory activities in the treatment group during a whole semester, while maintaining the same educational objectives, teaching program and organizational as well as temporal framework as the control group (sect. 3.3). In addition, a test for conceptual learning in mechanics, based on the prior research just mentioned, and aligned with the local study program was developed and re-validated on the basis of existing, well-validated tests (sect. 3.4.1). Furthermore, for assessing motivational outcomes, we utilized well-established instruments from the literature (sect. 3.4.2).

On the basis of the above research background and our own development for the intervention and instruments, our research questions are as follows: What are

1. the cognitive effects (conceptual learning, cognitive load) and
2. the motivational effects (interest, perceived relation to reality, self-concept and curiosity related to the studied topics)

of a MDET-based teaching sequence on mechanics for high school learners, as compared to a sequence without MDETs?

## 3 Methods

### 3.1 Sample and setting

The study was carried out on 16 to 17 years old learners of the Geneva high-school, having physics class-groups with a maximum size of 16 learners. A sample breakup is given in Tab. 2; 102 learners participated in the pilot study (PS) belonging to seven physics class-groups taught by three different teachers, and 111 learners participated in the main study (MS), belonging to eight class groups (four pairs) taught by four different teachers. In the main study, each teacher taught one control and one treatment class.

The pilot study was run to validate instruments and the feasibility of the intervention. In particular, a conceptual learning questionnaire, adapted to the specific learning content of the intervention and the local curriculum, was newly developed (see sect. 3.4.1), and several questionnaires for predictor and motivational variables were translated to French (see sect. 3.4.2, 3.4.3). An item analysis was carried out on theses questionnaires (see below). Given the fairly good progress and achievement of the pilot study, the main study could be conducted in the same way, with minor adaptations to the teaching program and in the MDETs activities. Pre- and post- questionnaires also worked well during the first study and could also be used up to minor adaptations.

Two further details have to be mentioned: first, as the main study took place during the school year 2019-2020, group 3 (see Tab. 2) had to take the post-tests remotely during the first week of confinement due to the Covid pandemic. Secondly, the learners of group 4 were in their 3$^{rd}$ year over a total of four years of upper secondary school, while the learners of the other groups were in their 2nd year. While the program of the mechanics course was exactly the same, these students had more mathematics and physics (on other topics) then the rest of the sample.

Tab. 2: Sample of the study. Each participating teacher taught at least one control and one treatment class-group. Sample sizes are given as total number ($N_{tot}$) / number of female ($N_f$) / number of male ($N_m$) participants; CG, TG: control and treatment group.

|  | TG | CG | TG + CG | Teachers | Class-groups |
|---|---|---|---|---|---|
|  | $N_{tot}/N_f/N_m$ | $N_{tot}/N_f/N_m$ | $N_{tot}/N_f/N_m$ | $N_{tot}/N_f/N_m$ |  |
| Pilot study (PS) | 59 / 39 / 20 | 43 / 23 / 20 | 102 / 62 / 40 | 3 / 2 / 1 | 7 |
| Main study (MS) | 55 / 33 / 22 | 56 / 31 / 25 | 111 / 64 / 47 | 4 / 2 / 2 | 8 |

### 3.2 Study and course design

The design was a repeated measurement control-treatment group comparison (control group vs. treatment group: without vs. with MDETs), which took place during learners' regular physics lessons. The intervention lasted a whole semester of 19 weeks. The teaching sequence followed the official teaching program[4], with the following content:

1) Introduction to kinematics: position vector, displacement, trajectory, velocity vector (average and instantaneous), speed (average and instantaneous), acceleration vector (mean and instantaneous);
2) One dimensional kinematics: uniform linear motion (ULM), uniformly accelerated linear motion (UALM), free fall;
3) Dynamics: Newton's laws 1, 2, 3 and the law of universal gravitation.

The course design is shown in Tab. 3. There were 13 weeks of classes, each consisting of two physics lessons, plus six weeks of testing and exams. In the treatment group, six activities based on MDETs replaced traditional experiments and exercises (i.e., there was the same number of experiments in both groups); the lab sessions had equal duration for control group and treatment group.

Tab. 3: Timeline of the sequence and tests. Standard tests refer to the assessment tests made by the teachers, independently on the study; they contribute to the final average grade of physics.

| Week | Part | Treatment group | Control group |
|---|---|---|---|
| 1 |  | Pre-tests ||
| 2 to 5 | 1 | Common lesson structure: introduction on physics background, paper-pencil exercises, corrections/answers for exercises and questions ||
|  |  | MDET lab sessions and exercises | Conventional lab sessions and exercises |
| 6 |  | Standard test ||
| 7 to 10 | 2 | Schedule identical to chapter 1 ||
| 11 |  | Standard test ||
| 12 | 3 | Schedule identical to chapter 1 ||
| 13 and 14 |  | Exam session ||
| 15 to 18 | 4 | Schedule identical to chapter 1 ||
| 19 |  | Post-tests ||

Up to the lab activities, the lessons had a common structure (introduction for physics background, paper-pencil exercises, corrections/answers for exercises and questions) identical for control group and treatment group, and the learning content was also identical. Both groups were taught by the same four teachers. The precise order of these activities varied according to the teaching practices of each teacher, but the total proportion of time dedicated to each type of activity was the same for control group and treatment group across all class groups. Moreover, the same topical context was provided in both the control group and treatment group. When, for practical reasons, the conditions did not allow the control group to carry out an experiment with the same topical context as treatment group (see example below), a conventional experiment was carried and the topical context in question was presented in an exercise sheet. All the participating teachers successfully completed the sequence and the test sessions, both in the PS and in the MS.

---

[4] https://www.ge.ch/document/programmes-disciplines-enseignees-dans-filiere-gymnasiale-au-college-geneve-valable-eleves-scolarises-2021

### 3.3    Learning materials

The MDETs activities were designed and carried out by the researchers and the participating teachers, according to the educational objectives, the level of the learners and the constraints related to the curriculum and the school setting. Based on the experiences of the PS, a few adaptations to the chronology and protocols of the activities were made in the MS. Particular care was taken to align the educational objectives with the contents and time constraints set by the school programs as well as with conceptual obstacles for the learning domain known from research (see 3.4.1). In particular, although the apps directly provide velocities, motion graphs, and regression analysis, the use of these features was introduced gradually within the sequence of activities (see Appendix 2). Students only made use of them once they had performed the calculations and produced the graphs themselves, ensuring they had understood the meaning of all the physical quantities and the relationship between the observed motion, the associated functions, and their mathematical and graphical representation.

The activities were carried out using eight iPads provided to each treatment class-group, in order to have at least one tablet for a pair of learners during the activity sessions. The applications "Video Physics" and "Graphical Analysis" by Vernier were used[5].

An example of a MDETs labwork is shown in Fig. 1, the standing vertical jump [4]. The process is divided into two main phases: The first one is the stand-up (acceleration) phase, when the feet are still in contact with the ground (between $t_0$ and $t_1$ in Fig. 1). Here the acceleration and the resultant net force are both constant and upward. The constant acceleration, both in the first and in the second phase, is found calculating the slope of the velocity-to time graph. The learners could draw the force diagram and had to apply their knowledge of dynamics (gravitation and the Newton's 2nd and 3rd laws) in order to find the pushing force of the person on the ground. The second phase (between $t_1$ and $t_2$ in Fig. 1) is the free motion under gravity (first upward, then downwards) with a change in the orientation of the velocity, starting when the feet are no longer in contact with the ground. This kind of motion is central in the mechanics curriculum, as it allows the learner to overcome the difficulty of differentiating velocity (which changes orientation during movement) and acceleration (which is constant and opposed to initial velocity). Moreover, the link between the acceleration and the resulting force is also established in the experiment.

More examples of MDETs activities are provided in Appendix 2, a detailed description of all the activities, protocols and worksheets is available in [73]. The following comment about the selection of MDETs activities is in order. We emphasize that a main purpose of our work was to implement and study MDETs in the real classroom setting of a regular high school mechanics course, with all its curricular, material, and practical constraints. This implies limitations on the kinds of the student experiments that can be carried out and the topical contexts that can be realized (e.g., no amusement park physics, no project labs — at least not in the local curriculum, etc.). Nevertheless, the material context was combined with a topical context wherever conditions allowed it (in particular the "round trip" and "jump" activity, see below); moreover, the aspect of "ownership of data" is present in all activities.

---

[5] https://www.vernier.com/product/video-physics-for-ios/

### 3.4 Variables and instruments

In this section, we will present the choice of variables and the construction of instruments. A more detailed description of the instruments' construction process as well as all the questionnaires and tests administered are available in the reference [73].

#### 3.4.1 *Conceptual Learning*

Conceptual learning was assessed pre- and post-test using a multiple-choice questionnaire adapted from the Force Concept Inventory (FCI [76]), the Test of Understanding Graphs in Kinematics (TUG [77]) and the Motion Concept Test, MCT [78]. Analogous to the sequence's learning activities, the concept learning test was designed within the curriculum, aligned with the research questions, and targeting common misconceptions. Each item was taken from of based on instruments already existing and validated in the literature. When no items addressed specific concepts, new questions aligned with the sequence's learning content were created and validated before the PS. The test of the PS was subsequently adapted and refined for the MS. The majority of the items were about one-dimensional kinematics, which represents most of the program covered during the duration of the study. Among others, questions about the interpretation of a stroboscopic image of a movement were included. These interpretations were particularly relevant in this study, because the tracing of a motion by the MDETs' apps provides exactly this kind of real-time representations, and therefore a difference in the results between the control group and the treatment group could be expected. Then, items testing misconceptions on vector displacement, average velocity and average acceleration in a two-dimensional motion, on free fall, on average speed in a round trip, on the difference between velocity and acceleration and on the three Newton's laws were included. The concept instrument of the MS consisted of 16 items in the pre-test and 19 items in the post-test, scored with 1 point for a unique right answer. The total score was the total ratio of right answers. The result of the conceptual learning post-test (LPO) was included in the calculation of the learners' final physics grade, whereas the result of the conceptual learning pre-test (LPR) was not. In order to motivate learners to apply themselves to sincerely answer the pre-test questions, they were explained that the results of these tests would be important to assess their progress in the course in view of a better implementation of teaching.

#### 3.4.2 *Motivational variables*

We studied the impact of the use of MDETs on four motivational variables, interest, perceived relation to reality, self-concept, and curiosity state. Instruments (six-level Likert scales) were based on well-validated instruments ([79] [80] [81] [82] [83], further developed and validated in [19] and [32]). After translation to French and re-validation in the pilot study, each final subscale contained between 5 and 7 items (on a 6- level Likert scale; 1 = completely disagree, 6 = completely agree). Sample items and results of the re-validation are reported in section 4.2.

#### 3.4.3 *Predictor variables*

A set of potentially influential predictors with well-validated instruments is available from prior research in the field [19] [27] [31]. We summarize key features for predictors for cognitive and affective outcomes: rationale, sample items and original sources, where more details can be found (abbreviations for later use in tables etc. are also included). In addition to the pre-values of each outcome variable, six predictors related to learner characteristics were taken into account:

- *Prior physics grade* (PY_GRADE_PRE) and *prior mathematics grade* (MA_GRADE_PRE), which serve as indicators of initial general knowledge in the two disciplines
- *Prior grade in in the teaching language* (French), since all the activities, exercises and the content

of the physics course are based on the understanding of spoken and written language (FR_GRADE_PRE).

- *Spatial abilities* (SA), assessed by the "paper folding" test well validated from previous research ([84], 176pp), and discussed as important factor for science learning with multiple representations [85], in particular in abstract domains like physics [86].
- *Curiosity as a trait* (CT), based on [82] and [83]; serves as baseline for curiosity state (3.4.2), one of the motivational outcomes. Sample item: "I find it fascinating to learn new things".
- *Self-concept regarding the use of smartphone apps* (SCS; based on and adapted from [87] [88] [89] and [19]), because individuals with a positive self-concept might show higher motivation and performance in learning with smartphones. Sample item: "I am good using the apps for Smartphones."

Five more predictors related to characteristics of learning and classroom processes were considered:

- *Cognitive load related to the experiments* and t*o the use of the app for Smartphones* (CLE and CLS, based on [90] [91] [92] and validated by [93] [94] [95]), as high CL might be harmful for both learning and motivation (sect. 2.2). Sample items: "I could well focus on the experiments, without struggling with the equipment" (CLE) and "I could well focus on the experiments, without struggling with the apps for the smartphone" (CLS).
- *Cognitive activation during the experiments* (CAE; based on [96]), and
- *Involvement during the physics lessons* (INV), as perceived by students; based on and adapted from [97]; [98] (WIHIC). Both CAE and INV belong to main positive mechanisms hypothesized for context-based science education (see sect. 2.1). Sample items: "I was actively involved in doing the experiments" (CAE) and "I expressed my thoughts during class discussions" (INV).
- Assessment (by students) of the *teacher engagement* (TA), also based and adapted from [97] [98] (WIHIC) in order to control for potential teacher difference between the treatment groups. Sample item: "The teacher's explanations helped me to understand".

All but two of the above instruments were used and validated in prior research on MDETs use in physics labs [19] [31]. Within the available testing time (45' in total) the two additional tests which could be included are those for *cognitive activation during the experiments and involvement during the physics lesson*, also well validated by prior research (see above). All tests had had satisfactory to good psychometric properties (e.g. all internal consistencies $\alpha_C \geq 0.75$, see [19] [96] [97]). After translation to French and re-validation in the pilot study, final subscales contained 5 to 7 items (on a 6-level Likert scale; 1 = completely disagree, 6 = completely agree). Results of the re-validation are reported in 4.2.

Additionally, analysis were carried out for gender and for different class-groups (i.e. a certain class taught by a certain teacher). For analysis of this set of predictors see 3.5.2.

## 3.5 Data analysis
### 3.5.1 *Analysis of items and instruments*

Test characteristics (discrimination *D*, item-test correlation $r_{it}$, and internal consistency, calculated as Cronbach's alpha $\alpha_C$) were calculated and item and instrument analysis were conducted out according to standard procedures and requirements [99] [100]. We also calculate for each item $\alpha^*$, the value of alpha if the item is removed, which can be used for scale improvement procedures [101].

For simplified comparison and interpretation, we follow Cohen et al. [102] who suggest a linear transformation of raw scores to the fraction (or percentage) of the possible maximum score. The values of psychometric scales for motivational variables and for the conceptual and spatial abilities tests were thus transformed to the interval between 0 and 1, according to $(S_o − S_{min})/(S_{max} − S_{min})$, where $S_o$ is a given observed score, and $S_{min}$ and $S_{max}$ are the minimum and maximum possible scores for a given scale, respectively.

Learning mechanics includes several separate but related concepts, such as interpreting one-dimensional or two-dimensional graphs, understanding the Newton's laws, free fall, etc. In order to test for potential sub-dimensions of the test of conceptual understanding, a factor analysis was additionally carried out ([103]; details of the procedure are reported in the appendix).

### 3.5.2 *Analysis of motivational and learning effects*

In order to test for effects of the intervention on motivational and learning outcomes in presence of predictors, analysis of (co)variance (AN(C)OVA) were carried out [104] [105]. First, it was ensured that the underlying assumptions are respected (independence and Gaussian distribution of residuals; homogeneity of residuals' variances; and, for analysis of covariance, homogeneity of regression slopes [105]). Second, in order to test for possible influences of predictors, we follow the approach recommended in [103] and [106], combining an a-priori selection of predictors available in the literature (see 3.4.3) with a preliminary statistical test for potentially significant influences (ANOVA for a given combination of outcome and predictor variable). Finally, once all significant predictors covariates were determined, we took them into account in the model and proceeded with the main analysis (ANCOVA). A correction for false discoveries was applied using the Benjamini-Hochberg procedure for multiple comparisons [107].

For comparing changes before and after the instruction or between the treatment group and the, effect sizes are reported as Cohen's *d* [108]. Usual effect-size levels (as established from comparison of a great many studies in different areas) are small ($0.2 < d < 0.5$), medium ($0.5 \leq d < 0.8$), and large ($0.8 \leq d$) [108]. Another reference value of $d = 0.4$ is used by Hattie [109] as a "hinge point" between influences of smaller and larger size.[6] Effects sizes when comparing more than two groups (after an ANCOVA) are reported as total eta squared, $\eta_t^2$ [108]. Effect-size levels for $\eta_t^2$ are small ($0.01 \leq \eta_t^2 < 0.06$), medium ($0.06 \leq \eta_t^2 < 0.14$), and large ($0.14 \leq \eta_t^2$); the same caveat[6] as for the *d* thresholds applies.

Analysis were carried out using the R software package [110] and Excel® and the data are available in [73].

## 4 Results I: Instrument validation

In this section we present an overview of the main results of the item and test analysis for learning and motivational outcomes, as well as for predictor variables. All results refer to the post-test of the main study, if not stated otherwise. Detailed results of the items and instruments of all tests (pilot study, pre-tests) can be found in the reference [73].

### 4.1 *Learning test*

The newly developed conceptual learning questionnaire, adapted to the specific learning content of the intervention and the local curriculum (see sect. 3.4.1), was analyzed according to standard procedures and requirements ([99] p. 16; see sect. 3.5.1). The conceptual test was constructed based on several a priori hypnotized, potentially interrelated conceptual dimensions: one-dimensional kinematics; free fall; Newton 1st, 2nd and 3rd law. The interdependence and, at the same time, the diversity of the items of the conceptual test was the consequence of the educational objectives discussed in section 3.2, and is a current property of classroom tests [111] [112]. In order to look for potential sub-dimensions of the test, an exploratory factor analysis was carried out. It revealed only one factor with 12 items, related to rectilinear motion, and no other factors composed of more than 3 items. Tab. 4 shows the results of the item analysis for both the full and restricted test (12 and 19 items, respectively).

---

[6] We agree with Hattie [139] that these thresholds values are an element of discussion to be used with circumspection, not a criterion to be applied blindly. Lower effect sizes might well be worth considering, depending on available alternatives, effort, and so on, and vice versa for higher effect sizes.

Tab. 4: Instrument analysis of the conceptual post-test of the main study for the full and restricted version (19 and 12 items, respectively; see text (N = 111; M = mean; SD = standard deviation; CI = confidence interval, for $\alpha_C$).

| Characteristics | Full version M(SD/CI) | Restricted version | |
|---|---|---|---|
| | | M(SD) | range |
| Item difficulty $P$ | 0.43(0.19) | 0.49(0.17) | [0.20; 0.76] |
| Item discrimination $D$ | 0.41(0.22) | 0.56(0.13) | [0.37; 0.75] |
| Item-test correlation $r_{it}$ | 0.34(0.14) | 0.45(0.08) | [0.35; 0.57] |
| Internal consistency $\alpha_C$ | 0.59 (0.11) | 0.68(0.09) | – |

A comment on the values of $\alpha_C$ is in order: as emphasized by current practice and recommendations, assessment tools covering multiple concepts, especially in classroom settings, are expected to have low values of alpha, and that low internal consistency is not a significant barrier to the use of a test with meaningful content coverage [104] [113]. In their review on assessment instruments in science education, Adams and Wieman [111] also indicate that instruments designed to measure multiple concepts in a short time, such as those administered in typical classroom settings, may have low values of internal consistency. According to van Blerkom [114], typical classroom tests indeed show a range of values between 0.60 and 0.80. Therefore, a value around 0.6 falls in the acceptable range for group level measurements [115] [116]. This reasoning is consistent with the results for the single dimension revealed by factor analysis, related to the conceptual understanding of rectilinear motion. As Tab. 4 shows, the increased uniformity of the restricted tests leads to a larger internal consistency (as well as a larger item-test correlation).

In view of the above, we kept the full test (19 items) for further analysis as a reasonable compromise between content coverage and psychometric quality.

Tab. 5: Sample items and instrument analysis of the motivational variables (interest (IN), perceived relation to reality (RR), self-concept (SC), and curiosity state (CS)) at post-test ($\alpha_C$: Cronbach alpha; $r_{it}$: item-test correlation; SD: standard deviation; CI: confidence interval).

| Variable: example of item (Translated from French) | $\alpha_C$ (CI) | Mean $r_{it}$ (SD) | Range of values $r_{it}$ |
|---|---|---|---|
| IN: I liked the physics class. | 0.71(0.09) | 0.69(0.25) | [0.56; 0.76] |
| RR: The subjects of the physics course are useful for everyday situations. | 0.90(0.03) | 0.85(0.15) | [0.79; 0.90] |
| SC: My classmates thought I was good at physics. | 0.89(0.03) | 0.84(0.19) | [0.75; 0.89] |
| CS: The course aroused my curiosity about the topics covered. | 0.89(0.03) | 0.83(0.12) | [0.81; 0.86] |

### 4.2 Motivational variables

As in the case of the learning test, instruments for the four motivational variables (interest, perceived relation to reality, self-concept, and curiosity state) were analyzed according to standard procedures and requirements [99] [117]. All test properties were within the recommended ranges, attaining satisfactory to good values [99]; see Tab. 5 for post-test values (values at pre-test are very similar).

### 4.3 Predictor variables

As for the dependent variables, the predictor variable instruments were also analyzed according to standard procedures and requirements [99] [117]. Results are reported in Tab. 5 and showed satisfactory values.

# 5 Results II: Intervention effects

In this section we present the main results obtained on the whole sample participating in the main study.

## 5.1 Learning

Tab. 6 reports descriptive values of the concept test before and after the instruction, specified for class groups (G1-G4: pairs of class groups taught by different teachers), control group and treatment group, as well as for all participants, together with values for Cohen's *d*. The results of the paired t-tests revealed significant gains for learning, with large effects sizes ($d > 0.8$) for all but one groups (we will come back to the one group with a medium size effect below). Thus, the instruction was globally effective, both for treatment group and control group (it removed successfully most of the preexisting misconceptions: the number of the most frequent incorrect answers fell from 60% in the pre-test to 25% in the post-test).

Tab. 6: Descriptive values for *learning pre and post-test* (LPR, LPO) and t-test statistics for the different groups (CG, TG: control and treatment group; G1-G4: pairs of class groups taught by different teachers): Mean success rates *P* ("difficulties", [99]), standard deviations, *t*-values, *p*-values, and Cohen's d for pre/post gains.

| Group | N | P (SD) LPR | P (SD) LPO | t | p | Cohen's d |
|---|---|---|---|---|---|---|
| **G1** | 27 | 0.28(0.15) | 0.49(0.14) | 5.03 | <0.001 | 1.4 |
| **G2** | 26 | 0.27(0.08) | 0.43(0.16) | 4.70 | <0.001 | 1.3 |
| **G3** | 27 | 0.25(0.11) | 0.32(0.16) | 1.91 | <0.001 | 0.5 |
| **G4** | 31 | 0.36(0.14) | 0.49(0.14) | 3.78 | <0.001 | 0.9 |
| **TG** | 55 | 0.31(0.13) | 0.44(0.17) | 4.26 | <0.001 | 0.9 |
| **CG** | 56 | 0.27(0.13) | 0.43(0.15) | 5.95 | <0.001 | 1.1 |
| **Tot** | 111 | 0.29(0.13) | 0.43(0.16) | 7.17 | <0.001 | 1.0 |

Furthermore, the following findings of the ANCOVA about learning outcomes are noteworthy (see Tab. 6). No significant effect was observed of the independent variable *treatment* ($F(1, 109) = 0.026$, $p = 0.87$). A medium size effect of the variable *class group* (G1-G4) on the result of the *learning post-test* (LPO) was observed ($F(3, 96) = 5.2$, $p<0.01$, $\eta_t^2 = 0.10$), as well as a small effect of the predictor *learning pre-test* ($F(1, 96) = 8.6$, $p<0.01$, $\eta_t^2 = 0.06$). No other predictors turned out to be significant for the conceptual post-test, specifically, no significant effects were found for *gender,* cognitive variables (prior *grades in mathematics, physics* and *French, spatial abilities, cognitive load* and *activation* during the experiments, *cognitive load* using the apps, *involvement* during the lessons), nor affective variables (*relation to reality, self-concept* regarding physics or using the apps, and *curiosity state* and *trait*).[7]

An additional ANCOVA on the individual item level of the learning post-test was carried out. Taking into account the correction for multiple significance testing [107], no differences were found between the control and the treatment groups.

---

[7] As the predictor *teacher assessment* is highly correlated with the variable *class group*, we did not take it into account for the ANCOVA.

Tab. 7: Significant results of the ANCOVA analysis on the concept learning post-test (LPO: learning post-test; class group; LPR: learning pre-test). The */** indicates a small/medium effect.

| LPO | Sum Squared | Degrees of freedom | F-value | p | $\eta_t^2$ |
|---|---|---|---|---|---|
| Treatment | 7.45 | 1 | 1.18 | 0.28 | – |
| Class group (G1-G4) | 101.1 | 3 | 5.34 | <0.01 | 0.10** |
| LPR | 59.7 | 1 | 9.45 | <0.01 | 0.06* |
| Residuals | 600.0 | 95 | | | |

Moreover, an ANCOVA for the *post average grade in mathematics* was carried out, and we found an indication of a possible positive effect of the treatment ($F(1, 102) = 2.9$, $p = 0.09$, $\eta_t^2 = 0.03$). Although the *p*-value for this hypothetical effect is only marginally significant, it is worth considering the possibility of such an effect, as we will discuss in section 7.

### 5.2 Cognitive load, cognitive activation, and motivational variables

First, the mean values of the motivational and other control variables for all participants as well as for control group and treatment group are reported in Tab. 8. We observe that the values for all these variables are comparable for the control group and for the treatment group. In particular there was no difference in *cognitive load during the experiment* between the learning groups with and without MDETs ($t(107) = 0.48$; $p = 0.6$).

Tab. 8: Mean value (total / TG (treatment group) / CG (control group)) and standard deviation of cognitive load, cognitive activation, and motivational variables (CT: curiosity trait; CLE/CLS: cognitive load related to the experiments/ use of the smartphone app; CAE: cognitive activation during the experiments; INV: student involvement during the physics lessons; SCS: self-concept regarding use of smartphones/apps; TA: teacher assessment).

| Variable and sample item (translated from French) | Mean value (SD): total / TG / CG | range |
|---|---|---|
| CT: I find it fascinating to learn new things. | 4.62(0.71) / 4.72(0.73) / 4.53(0.69) | [3.79 ; 5.22] |
| CAE: I was actively involved in doing the experiments. | 4.16(0.73) / 4.07(0.73) / 4.24(0.74) | [3.84 ; 4.59] |
| INV: I shared my thoughts during the class discussions. | 3.60(0.95) / 3.56(0.93) / 3.63(0.99) | [3.19 ; 3.99] |
| SCS: I am good using the apps for Smartphone. | 4.48(0.88) / 4.40(0.84) / 4.55(0.92) | [4.19 ; 4.79] |
| CLE: I could well focus on the experiments, without struggling with the equipment. | 4.06(0.82) / 4.10(0.87) / 4.03(0.78) | [3.40 ; 4.68] |
| CLS: I could well focus on the experiments, without struggling with the apps for Smartphone. | Only TG: 4.35(1.15) | [3.78 ; 4.76] |
| TA: The teacher's explanations helped me to understand. | 4.91(0.43) / 4.88(0.56) / 4.93(0.27) | [4.46 ; 5.24] |

Second, descriptive values for the motivational dependent variables (*interest, perceived relation to reality, self-concept,* and *curiosity state*) before and after the instruction are reported in Tab. 9. The t-test results showed for the whole sample a significant gain for *perceived relation to reality* (RR, $t(219.57) = 2.2$, $p = 0.03$), with an effect size of $d = 0.29$. No significant changes were found for *interest* (IN, $t(219.93) = 1.6$, $p = 0.12$), *self-concept* (SC, $t(216.6) = 0.28$, $p = 0.77$) and *curiosity state* (CS, $t(215.87) = 0.60$, $p = 0.55$).

From the ANCOVA results, no significant effects of the treatment (control group or treatment group) on the motivational dependent variables were found ($F(1, 109) = 0.92$, $p = 0.34$ for *interest*; $F(1, 109) = 0.65$, $p = 0.42$ for *curiosity state*; $F(1, 109) = 0.009$ and $p = 0.92$ for *self-concept*; $F(1, 109) = 0.27$, $p = 0.61$ for *relation to reality)*. Tab. 10 reports the ANCOVA results for motivational outcomes, focusing on those variables, where significant pre/post changes have been found.

<u>Interest</u>: a medium effect of the variable *class group* (G1-G4) was found in the post-test, even when controlling for *involvement* and *cognitive activation* during the experiments ($F(3, 101) = 3.3$, $p = 0.02$ $\eta_t^2 = 0.04$, see Tab. 10). Note that no other predictors turned out to be significant for interest at post-test.

<u>Self-concept</u>: Significant effects (with small effects sizes) of the variables prior *self-concept* ($F(1,89) = 3.3$, $p < 0.01$, $\eta_t^2 = 0.05$), and *physics grade* ($F(1,89) = 3.1$, $p < 0.01$, $\eta_t^2 = 0.04)$, as well as of *gender* on *self-concept* were found, consistent with prior research [118]-[120]. No other predictors turned out to be significant for self-concept at post-test.

<u>Curiosity state</u>: There were significant small-to-medium effects of the predictors *prior curiosity state* ($F(1, 99) = 18.6$, $p < 0.01$, $\eta_t^2 = 0.09$), *cognitive activation* during the experiments ($F(1, 99) = 4.6$, $p = 0.04$, $\eta_t^2 = 0.02)$, and *involvement* ($F(1, 99) = 4.2$, $p = 0.04$, $\eta_t^2 = 0.02$), on the variable *curiosity state* at post-test. No other predictors were found to be significant for self- curiosity at post-test.

<u>Perceived relation to reality (POST_RR):</u> Only the pre-test value (PRE_RR) was found to have an effect on the output variable post *perceived relation to reality at post-test* (of medium size, $F(1, 101) = 14.0$, $p<0.01$, $\eta_t^2 = 0.08$).

Tab. 9: Descriptive values and t-test statistics of the pre- and post-tests for motivational variables: mean values (*M*), standard deviation (SD) and t- and p-values (6-level scale, 1 = completely disagree, 6 = completely agree), for TG (treatment group) / CG (control group) / total.

| | | Interest (IN) | | | | | | Curiosity state (CS) | | | |
|---|---|---|---|---|---|---|---|---|---|---|---|
| Group | N | M (SD) PRE | M (SD) POST | t | p | Group | N | M (SD) PRE | M(SD) POST | t | p |
| TG | 55 | 3.10(0.88) | 3.27(0.90) | 1.0 | 0.31 | TG | 55 | 3.40(1.04) | 3.34(1.05) | 0.29 | 0.77 |
| CG | 56 | 3.24(0.82) | 3.43(0.83) | 1.2 | 0.23 | CG | 56 | 3.28(0.85) | 3.17(1.13) | 0.56 | 0.57 |
| Tot | 111 | 3.17(0.85) | 3.35(0.87) | 1.6 | 0.12 | Tot | 111 | 3.34(0.95) | 3.25(1.09) | 0.60 | 0.55 |
| | | Self-concept (SC) | | | | | | Perceived relation to reality (RR) | | | |
| Group | N | M (SD) PRE | M (SD) POST | t | p | Group | N | M (SD) PRE | M (SD) POST | t | p |
| TG | 55 | 3.52(1.07) | 3.59(1.17) | 0.31 | 0.76 | TG | 55 | 3.44(1.04) | 3.73(1.00) | 1.4 | 0.15 |
| CG | 56 | 3.59(0.97) | 3.61(1.15) | 0.09 | 0.93 | CG | 56 | 3.25(1.12) | 3.61(1.25) | 1.6 | 0.11 |
| Tot | 111 | 3.56(1.02) | 3.60(1.15) | 0.28 | 0.77 | Tot | 111 | 3.35(1.08) | 3.67(1.13) | 2.2 | 0.03 |

Tab. 10: ANCOVA results for motivational outcome variables at post-test; */** indicate small/medium effects.
<u>Outcome variables</u> are: POST_IN (interest at post-test), POST_SC (self-concept at post-test), POST_CS (curiosity state at post-test). POST_RR (perceived relation to reality at post-test
<u>Predictors</u>: CAE ( cognitive activation related to the experiments), CLE (cognitive load related to the experiments), INV ( student involvement during the physics lessons), PRE_CS (curiosity state at pre-test), PRE_SC ( self-concept at pre-test), PRE_RR (perceived relation to reality at pre-test) and PY_GRADE_POST (physics average grade at post-test).

| POST_IN | Sum of Squares | Degrees of freedom | F-value | p | $\eta_t^2$ |
|---|---|---|---|---|---|
| class group (G1-G4) | 3.5 | 3 | 3.3 | 0.02 | 0.04* |
| CAE | 2.1 | 1 | 6.1 | 0.02 | 0.06** |
| INV | 5.1 | 1 | 14.4 | <0.01 | 0.12** |
| Residuals | 35.6 | 101 | | | |
| **POST_SC** | | | | | |
| PRE_SC | 3.3 | 1 | 9.3 | <0.01 | 0.05* |
| GENDER | 2.4 | 1 | 6.8 | 0.01 | 0.03* |
| PY_GRADE_POST | 3.1 | 1 | 8.6 | <0.01 | 0.04* |
| CLE | 1.9 | 1 | 5.3 | 0.03 | 0.03* |
| Residuals | 14.0 | 89 | | | |
| **POST_CS** | | | | | |
| PRE_CS | 12.4 | 1 | 18.6 | <0.01 | 0.09** |
| CAE | 3.0 | 1 | 4.6 | 0.04 | 0.02* |
| INV | 2.8 | 1 | 4.2 | 0.04 | 0.02* |
| Residuals | 65.9 | 99 | | | |
| **POST_RR** | | | | | |
| PRE_RR | 11.7 | 1 | 14.0 | <0.01 | 0.08** |
| CAE | 2.5 | 1 | 3.0 | 0.09 | n.s. |
| CLE | 2.5 | 1 | 3.0 | 0.09 | n.s. |
| Residuals | 84.1 | 101 | | | |

# 6 Discussion
## 6.1 Cognitive aspects

Our findings show that an overall learning effect occurred, with a pre/post effect size (Cohen *d*) around 1, indicating a large effect. The beneficial effect of the instruction on conceptual learning also occurred separately in all individual class-groups. However, although an overall strong learning gain of conceptual understanding of mechanics was observed both for control group and treatment group, the answer to our first research question is that there is no significant difference in physics learning between the control group and the treatment group. This result differs from those of the studies by Hochberg *et al*. [31] and Klein *et al*. [27], discussed above. A possible reason could be the differences in the target groups of these studies. Here we examined high school learners of non-specialized physics classes, while the learners for whom the positive effects had been found in Hochberg *et al*. [31] were from high school specialized physics classes, and in Klein *et al*. [27] were first year physics university students. In both cases, these are learners with advanced physics understanding, and this might be a prerequisite for fully exploiting the cognitive potential of MDETs. Another factor for further improvement may be due to the setting of regular mechanics course with its tight constraints, limiting feasible experiments and topical contexts (see sect. 3.3).

On the other hand, similarly to what found by Hochberg *et al.* [19] [31], and Klein *et al.* [27], the use of MDETs did not increase the perceived cognitive load during the experiments. Moreover, the ANCOVA revealed no significant influence of perceived cognitive load due to the use of apps on the learning outcomes of the treatment group. Thus, no evidence of often apprehended disadvantages of over-challenging and distraction of learners by mobile devices was found in the present study.

Regarding features of successful labwork, another point beyond the use of MDETs is worth discussing. Recent research shows that traditional, highly structured undergraduate labs, reinforcing textbook content (e.g., verifying Hooke's law), do not enhance learning [121] [122]. Smith, Holmes et al. [121] [122] define such labs as combining content verification with rigid "cookbook" formats, leaving little room for autonomous student thought. We fully agree these formats are ineffective. While our lab course includes a verification, role focused on conceptual obstacles, its goals align well with those authors cited above (e.g. [123]). Specifically, our labs provide procedural guidance (though no "cookbook" protocols), aiming for content enhancement rather than mere reinforcement. They allow students to confront conceptual obstacles with hands-on, empirical evidence—a well-founded purpose of student experiments, especially at the high school level. With this clarification—that verification (or "enhancement") does not imply cookbook labs—we see a shared key factor with [121] and [122]: proper "active learning engagement" [124] [125] or "cognitive activation" [126] [127] for the given learning goals (in our case conceptual understanding). Our results confirm that MDETs labs can indeed provide such activation.

An effect of the different class groups (different classes taught by a different teacher) on the result of the conceptual test was observed (see sect. 5.1 and Tab. 7), with a medium effect size. This effect, mainly due the lower results of group 3 (see Tab. 6), can be explained by the fact the learners of this group completed the post-tests during the first Covid19 lockdown as a formative assessment, without influence on the final grade. This may have affected the learners' investment in learning both before and during the conceptual post-test.

A final observation is worth to be mentioned. While previous studies generally highlight the positive impact of mathematical understanding on physics learning [128-133], our study suggests a possible reverse effect of using the mathematics in a physics course context: the use of MDETs and of the cognitive effort required to understand multiple representations generated by them may enhance mathematics learning (see e.g. [134] and [135] for similar considerations).

### 6.2 Motivational aspects

Overall, the impact of the instruction was similar for the control group and the treatment group concerning motivational aspects. The pre/post variations of the perceived *relation to reality, self-concept, interest* and *curiosity state* in physics are similar for both groups and, in addition, they were all slightly positive. For *perceived relation to reality* this improvement was even significant for both treatment and control groups, with a small effect size ($d = 0.29$). These results showed that it was possible to maintain or even slightly improve the motivational aspects investigated in this study. Regarding the comparison between control group and treatment group, no significant differences were found. Thus, regarding our second research question on motivational aspects, a slight improvement could be established, but no superior effects by the MDETs treatment. In addition, we collected the participating teachers' perceptions on using mobile devices in physics labs, and none reported the observation of over-challenging or distracting effects on students' focus. They emphasized the practicality and ease of MDETs, which not only reduced repetitive tasks but also enabled experiments that would not be possible with traditional equipment. While their attitudes ranged from cautious to optimistic, all agreed on the value of the approach and expressed their intention to continue using MDETs in future classes.

## 6.3 Limitations

The following methodological limitations of the study have to be noted. First, there are interesting variations of the use of MDETs which could not be integrated in the design. One such possibility is using the learners' private Smartphones for the MDETs activities (Bring Your Own Device (BYOD; [136], [137]), which was not explored in this study, as in the Geneva schools the use of private smartphones during the lessons is prohibited. Moreover, this idea raises several additional issues, which would have strongly changed the intentions of this research. First, from a learning point of view, the distracting effects by private smartphones are likely to be stronger than those of devices provided by the institution, as the possibilities of personal use (images, music, games, communication) are present. Secondly, although smartphones and tablets are very widespread, some learners can afford models that perform better than others and some learners do not have any at all, therefore the question of equal treatment should also be considered. Furthermore, the apps used in the study are only available for a brand of high-end devices and there are no similar apps for all models of tablets or Smartphones. Asking the families of all learners to buy apps for their phone is problematic: families can be opposed, many learners do not have the right brand and, to date, no similar free-to-access video tracking apps is available. Thus, while the BYOD approach is are advocated by some authors, and indeed seems to have also promising features [138] [139] [140], it was not included in this study. A further discussion of pros and cons can be found in [141].

Second, there are limitations on the level of the variables taken into account in the present study. One such limitation is about cognitive load and the differentiation of its different types. For instance, high values of CL can be either positive or negative for learning, depending on the proportion of extraneous and germane cognitive load [142] [143]. Possibilities for empirically differentiating between the different types of CL exist [144], but were beyond the scope of this investigation.

Finally, the variable "group" (i.e. a certain class taught by a certain teacher) has a considerable influence on the outcomes for learning and interest. But the design and instruments of the present study do not allow for a more fine-graded analysis of the potential factors at work. This is a natural limitation of quasi-experimental investigations in real classroom settings, with their tight constraints e.g. on available testing time, and thus the number of variables which can be tested. Note, however, that in present study two measures to control for possible teacher influences were taken: the same teacher taught both a control and a treatment class, and additionally one dimension of the WIHIC classroom questionnaire [104] was taken into account, viz. teacher engagement (as perceived by students). While this allows for some control of overall (same teacher) and differential teacher effects (engagement questionnaire) on the comparison of treatments, a more detailed investigation would be of interest (see below).

A limitation of the present study is its setting within a regular mechanics course, which restricts the types of student experiments that can be implemented, thereby not fully exploiting the potential of topical contexts. Nevertheless, we see our work as a valuable contribution to expanding the research foundation for integrating MDETs into everyday instruction. Note also that, despite the above choices and their acknowledged limitations, the MDETs intervention still leads to strong pre/post gains in conceptual learning and even some gains in motivational outcomes. These findings, in turn, can serve as a basis for developing and studying more far-reaching ways of creating contexts with MDETs in more open teaching settings.

Finally, the following practical limitations should be considered when teaching with MDETs with high school students: first, there is a phase of appropriation of the mobile device (tablet or smartphone) and of the apps. It is generally not easy for learners to make a video of sufficient quality for the tracing by the app to be possible: is has to be stable during the recording, sufficient contrast between the studied object and the background is needed, the angular speed (pixel speed) cannot be too high, the tracing cursor has to be chosen with an appropriate width, etc. All these requirements take time to be learned and mastered, which may not be not evident for learners and, often, to teachers as well (one among the four participating teachers had expressed a certain degree of difficulty in that respect). Moreover, it's worth noticing that certain teachers who did not participate in the study did not even take into consideration to use these devices in their lessons, because they mainly affirmed that they could not master them with sufficient ease. This confirms what has emerged in previous investigations on the difficulties of integrating technologies in schools, including human factors or other barriers due to the technologies [145].

Second, beyond the practical mastery of the device, the use of MDETs may cause cognitive overload (see sect. 2.3), if learners do not yet master the concepts and mathematics underlying the representations provided by the apps and used during the experiments (e.g. diagrams of position or of velocity vs. time). On the one hand, it is well known that familiarizing with the different mathematical representations of a physical phenomenon requires an effort from learners [64] [142]. On the other hand, it might be this effort that could give a better knowledge of mathematics, see below. Thus, when planning a teaching sequence, teachers should carefully try to maintain a balance between productive efforts for learning and the risk of overloading the working memory. The present study shows that it is possible do so, but teachers should be aware that this requires special attention.

Third, despite initial intentions, it was not possible to let the learners carry out MDETs activities as homework, mainly for practical reasons. They would have had to take the devices home, which was not manageable from an organizational point of view for the limited availability of these devices (several classes in parallel used the same set of tablets furnished) and for the security of their use.

Finally, difficulties may arise in the support of tablets by schools, taking into account the maintenance and the technical assistance they request. It may seem trivial, but for example having a wi-fi network or a Bluetooth printer in each classroom is not obvious. Updates are constantly needed, and these high-tech objects become obsolete in few years, faster than standard lab equipment like rails, timers or dynamometers. Furthermore, we cannot neglect the planned obsolescence of these devices and their cost in environmental and durability terms [147]. All these factors should be put on a balance with hindsight, taking into account the global context and the fact that we imagine a scientific education sustainable and accessible to all the community.

# 7   Perspectives and Conclusions

## 7.1   Perspectives for classroom practice and further research

We begin with a series of observations regarding classroom practice. Taking into account its beneficial effects as well as the limitations described above, the use of MDETs turns out to be advisable under certain conditions, not as a replacement, but as a complement of traditional settings. For some course contents, it is possible to choose whether to do an experiment with traditional instruments or with MDETs (as for linear motions), while for other contents either the traditional activity (e.g. free fall, where the temporal resolution of the MDETs app is insufficient) or the MDETs activity (e.g. the "jump" activity, see sect. 3.3 and [4]) are clearly more advisable.

When possible, the easiness and practicality of the preparation of the experiments by the teacher constitutes a clear advantage: for the same content as in a conventional activity, MDETs activities allow a faster implementation with a handy and light device, allowing to save precious teaching time for other objectives. This advantage was highly appreciated by teachers participating in this study.

Teachers also appreciated the flexibility of MDETs in different classroom settings: depending on the class-group and the lesson plan, learners can take measurements and analyze them (experimental activity), or the data can be given from the measurement by a single student or by the teacher and used by the whole class (semi-experimental activity). In the latter case, the data ownership and the real-time data capture are not present, but the possibility of learning in authentic situations might be even enhanced, as the video may be recorded outside the classroom (playground, amusement park, etc.).

The following perspectives appear as interesting for both research and practice. First, the possibility of "home experiments". Learners can carry out activities at home, for example as homework, or when the lessons are given at a distance. This possibility has been considered in science education since almost two decades [148], and it has attracted new attention during the COVID-19 period, and beyond [149]. Educational institutions worldwide were obliged to adapt to remote learning [150] [151], and MDETs have shown to be a promising element in these strategies [152]. The insights learnt from these experiences continue to improve physics instruction. An empirical investigation of these ideas was not in the scope but would be of interest both for practice and research in science education.

Second, we mention the potential of MDETs activities to create for closer interaction between physics and mathematics education [153]. For example, mathematics classes could use representations obtained from physics experiments to illustrate how mathematical concepts apply to real-world contexts. Conversely, in physics courses, MDETs can support an enhanced integration of mathematics. An empirical study on the effects of MDETs on students' learning of mathematics as applied to physics thus appears promising.

Finally, the following aspects are also of interest for further research. We already mentioned that the differentiation of different types of cognitive load [144], potentially coupled to measures of short-term memory capacity, appears to be relevant for future research on the implications of the use of MDETs in the learning of physics. Another topic worth exploring is how to improve the understanding of the `group´ factor (a given combination of teacher and class) in terms of operationalizable variables. Such variables could be on a general level, such as assessed with one of the instruments of the WIHIC questionnaire (What Is Happening In this Class [104] [154] [155]), e.g. the dimensions of "Investigation" and "Task Orientation". On a more specific level, attitudes and competences of teachers regarding the use of technology in the science classroom are of interest, such as analyzed in the TPCK framework (Technological pedagogical content knowledge [156]).

## 7.2 Summary and Conclusions

In this research, we examined the impact of integrating Mobile Devices as Experimental Tools (MDETs) in physics laboratory sessions over one semester. In particular, we investigated how this integration influenced situational interest, curiosity, relation to reality, self-concept, conceptual learning in physics, and the acquisition of related mathematics among students enrolled in a non-specialized mechanics course. Among our findings, and we would like to highlight the following main points.

First, this is – to the best of our knowledge – the first empirical study about the educational effects of MDETs in a full teaching sequence on physics at the high school level. Specifically, the present contribution presents such an investigation in a mechanics high school course, with a quasi-experimental pre/post design comparing a treatment group with MDETs vs. a control group without, and taking into account various control measures (same content, lesson plan and teacher for control and treatment group; use of series of predictor variables). For the global learning gain, a statistically significant effect with an effect size $d = 0.9$ was found, which is quite a large effect. Compared to the control group, no significant difference was found. This appears as unexpected in view the theoretical advantages of the approach (sect. 2); however, as the learning gains of both control and treatment groups are quite large, one is dealing with a quite effective teaching in both cases. Possible reasons for the absence of a greater learning gain of the MDETs group are discussed. Moreover, evidence for "null effects" for new treatments, especially when accompanied with a certain "hype" of expectations as it is often the case with instructional technology, is considered as increasingly important in educational, social, health, and other sciences [157] [158], including physics education research [159]. Second, the present study did not find evidence supporting commonly anticipated drawbacks of instructional technology. Specifically, there was no difference regarding cognitive load between the control group and the treatment group. This suggests that when the use of mobile digital educational technologies (MDETs) is appropriately planned and guided by a teacher, they do not lead to an over-challenging of students. Moreover, the presence of positive learning gains and the absence of negative effects did not depend on various potential influence factors; in particular there was no or little dependence on gender, prior knowledge and other predictors. Put positively, the findings of this study can be summarized by saying that students can learn by the MDETs approach as much as by well-prepared highly effective conventional lessons (but not more), and that learners of both genders and with different initial conditions can have comparable learning gains (of course, the absolute level depends on the initial state).

Finally, MDETs thus appear as a useful extension of the toolkit of the physics teacher, comparable to other existing successful teaching practices, while offering a number of additional advantages (variation of teaching approaches, inclusion of digital skills, etc.) and potential further perspectives for research and development in physics education (extensions to non-standard physics experiments, e.g. as homework; synergies between physics and mathematics learning).

**Acknowledgements**: Swissuniversities has supported this project through its program on discipline-based education research. The implementation of the teaching materials and the study as a whole was only possible thanks to the close collaboration and very good mutual understanding between researchers, teachers, and school laboratory assistants throughout the process. Moreover, valuable technical support by Dr. L. Darmendrail (now Valdivia, Chile) and helpful advice by Dr. L. Weiss (Geneva) is gratefully acknowledged.

**Authors' Contributions:** Conceptualization: A.G, A.M; Formal Analysis: A.G., M.D, F.S.; Investigation: A. G, A.M; Methodology: A.G., M.D, A.M., F.S.; Project Administration: A. G, A.M; Supervision: A.M.; Writing: A.G, A.M, with contribution by M.D, F.S

# Appendices

**Appendix 1: Factor analysis**

Exploratory factor analysis was performed for the conceptual test, in view of the complexity of the concepts' interdependence in the targeted physics curriculum, and the absence of previous empirical evidence. The number of factors was determined by considering only the factors with eigenvalues > 1 [103]. The resulting subject-to-item ratio was about 6:1 (N = 111, k = 19 items), which can be considered as sufficient [160]. In order to identify the most suitable factor structure, we compared several solutions with different numbers of factors, used oblique rotation to simplify the data [160], and selected the solution that best aligned with theoretical and/or educational considerations based on a series of "goodness of fit" indices [161] [162]. Following [162], we considered that an item belonged to a given factor when its loading was greater than 0.3, unless a theoretical argument supported its presence.

**Appendix 2: Examples of MDETs activities**

The round trip: This activity consists of analyzing a return trip movement of a student (see Fig. 2). Each trip covers the same distance but is made at a faster speed on the outward journey than on the return. The learners have to calculate the velocity vector and the speed for the outward journey, for the return, and for the entire trip, after having made their prediction of the total average speed. This activity therefore allows the learners to confront the misconception that the total average speed is equal to the mathematical average of the speeds in each of the journeys of the same distance: according to the studies of Reed *et al*. [73] [74], 85% of the students give this answer and only 5% of the students give the right answer. For this activity, it was possible to keep the same related-to-reality situation for the control group and the treatment group: the learners of the control group could do the exact same activity, but measuring times and distances with traditional instruments, such as stopwatches or meters.

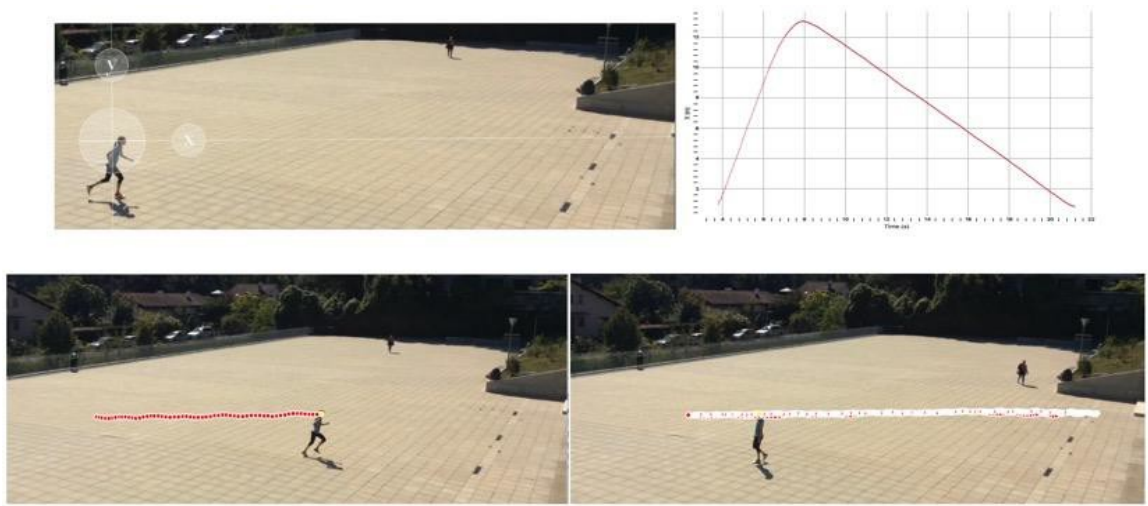

Fig. 2: Example of situation of round trip MDETs activity: fixing the reference frame, time diagram and tracing the movement.

Uniform Linear Motion (ULM): To analyze ULM, control group learners performed an experiment using a traditional setup, which varied according to the teacher and the school: for example, the motion of a ball over a short distance, of a marble or a cart on a rail (air rail or slightly inclined rail, to compensate for friction). The learners of the control group had to create the motion's time diagram from the data collected with traditional instruments like a chronometer and a measuring tape. The learners of treatment group analyzed the movement of an overhead projector film roll on a table (see

Fig. 3), starting from the data table provided by the app. As for the control group, they calculated the slope and the intercept, and they found that the diagram of velocity is a plateau. Indeed, they also got the graphics directly available by the app and thus didn't have to draw it. As one of the pedagogical objectives of this activity was constructing a time-diagram of a motion, learners of the treatment group did not use the option "add a regression" to the graph here initially. They were shown this function only at the end of their analysis, to compare with their results for the slope and the intercept.

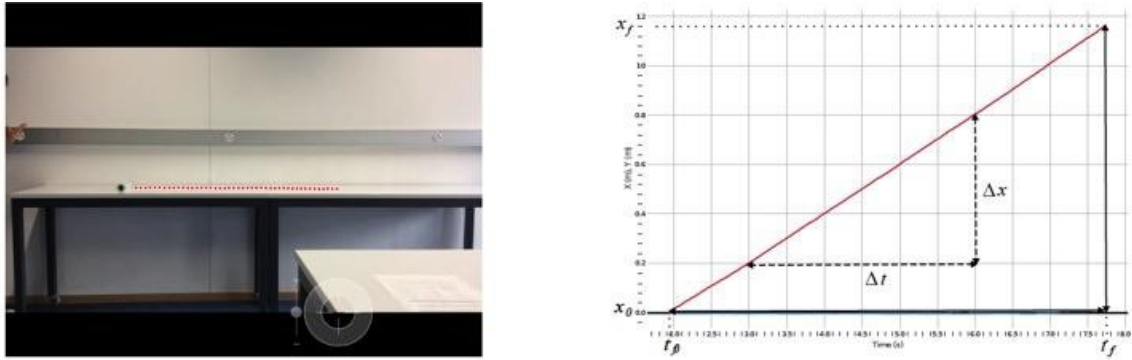

Fig. 3: The ULM of an overhead projector film roll: the chronophotography and the time diagram for the same motion.

For both groups, this activity was the first allowing the learners (i) to work on the notion of function and dependent and independent variables as well as (ii) to associate these abstract concepts with the corresponding physical quantities. Although the learners had seen these concepts before in math classes, this topic remains challenging for them.

Uniformly Accelerated Linear Motion (UALM): For the study of this motion type, both treatment and control groups worked with the classroom classical setup (inclined rail, see Fig. 4 on the left) and produced graphs of velocity (linear) and of position (quadratic) versus time for this motion (see the graphs in Fig. 4). As for the ULM activity, the control group learners had to produce the graphics reporting the data, while the treatment group learners used the graphics provided by the apps. Moreover, a regression curve can easily be added, so that the treatment group learners could better focus on the taught physics contents.

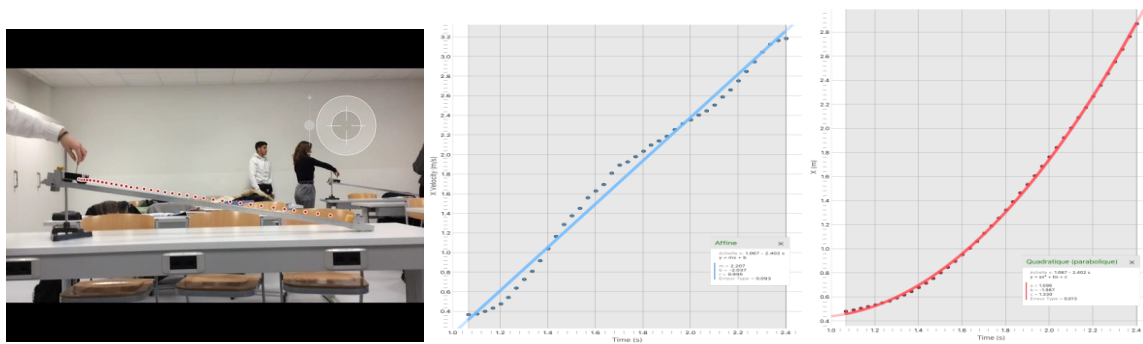

Fig. 4: Example of a MDETs activity on UALM including chronophotography, position-time and velocity-time graphs.

Projectile motion: In this activity learners of the treatment group traced the motion of a ball falling to the floor after rolling on a table (i.e. with an initial horizontal velocity). They first performed an analysis of the ULM ($x$-axis) and UALM ($y$-axis) by calculating the velocity or the acceleration from the graph given by the application, as shown in the example in Fig. 5. Subsequently, learners repeated the tracing for different initial speeds. Each time they measured the horizontal distance traveled by the ball during the flight and they established a proportionality relation between this distance and the initial speed. This allowed realizing that the fall time does not depend on the initial velocity but only on the height of the table.

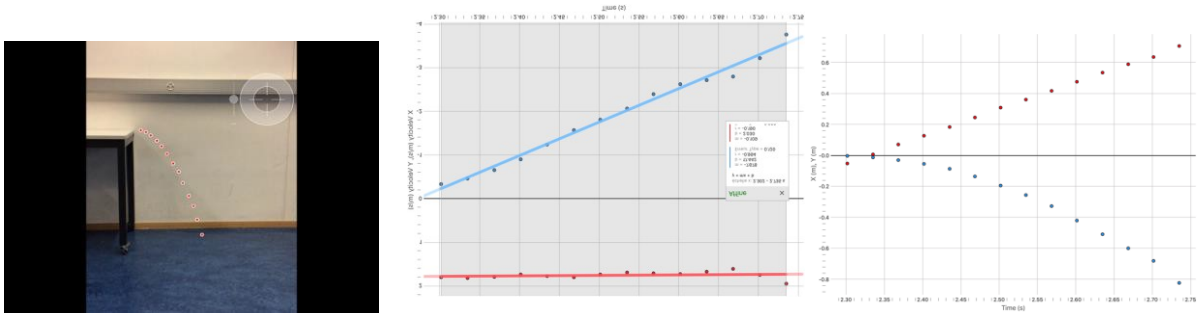

Fig. 5: Tracing, velocity and position-to-time diagrams outcomes of the apps for the projectile motion activity.

The equivalent traditional lab could be carried out by the control group using a stopwatch chronometer and a ruler to measure the ball's initial velocity (from its diameter and passing time) and the horizontal distance from the table, when the ball hit the ground.